\documentclass[prc,aps,12pt,final,notitlepage,oneside,twocolumn,
nobibnotes,nofootinbib,superscriptaddress,noshowpacs]{revtex4-1}
\usepackage{indentfirst}
\usepackage{epsfig,psfrag}
\usepackage{graphicx}
\usepackage{mathrsfs}
\usepackage{mathrsfs,amssymb}
\usepackage{natbib}
\usepackage{color}
\usepackage{ulem}
\usepackage[english]{babel}
\usepackage{bm}
\usepackage{color,graphicx}
\usepackage{graphicx,colordvi}
\usepackage{array}
\usepackage{calc}
\usepackage{ifthen}
\usepackage{epsfig}
\def\be{\begin{equation}}
\def\ee{\end{equation}}
\def\bea{\begin{eqnarray}}
\def\eea{\end{eqnarray}}
\def\l{\label}
\def\r{{\bf r}}
\def\p{{\bf p}}
\def\s{{\bf s}}

\def\om{\omega}

\def\sinh{\hbox{sinh}}

\def\exp{\hbox{exp}}

\def\sinh{\hbox{sinh}}

\def\Re{{\mbox {\rm Re}}}
\def\d{\hbox{d}}
\def\eps{\varepsilon}
\def\siml{\hbox{\kern.1em \lower.6ex \hbox{$\sim$} \kern-1.12em
 \raise.6ex \hbox{$<$} \kern.1em}}
\def\simg{\hbox{\kern.1em \lower.6ex \hbox{$\sim$} \kern-1.12em
 \raise.6ex \hbox{$>$} \kern.1em}}
\def\bs{\bigskip}
\def\ms{\medskip}
\begin{document}

\title{SURFACE CORRECTIONS TO THE MOMENT OF INERTIA AND
 SHELL STRUCTURE IN FINITE FERMI SYSTEMS}

\author{D.V. Gorpinchenko$^{\;1}$, A.G. Magner}
\affiliation{\it  Institute for Nuclear Research, 03680 Kyiv, Ukraine} 
\author {J. Bartel}
\affiliation{\it Universit\'e de Strasbourg, IPHC, F-67037 Strasbourg, France} 
\author {J.P. Blocki} 
\affiliation{\it National Centre for Nuclear Research,  PL-00681 
Warsaw, Poland}

\bigskip

\centerline{\today}
\ms

\maketitle

\vspace*{-1.2cm}

 \begin{center}
  \Large{\bf Abstract}
 \end{center}


The moment of inertia for nuclear collective rotations is derived  
within a semiclassical approach based on the Inglis cranking and
Strutinsky shell-correction methods, improved by  
surface corrections within 
the nonperturbative periodic-orbit theory. 
For adiabatic 
(statistical-equilibrium) rotations it was approximated by
the generalized rigid-body moment of inertia 
accounting for the shell corrections of the
particle density. An improved 
phase-space trace formula  allows 
to express the shell components of the moment of inertia  
more accurately in terms of the free-energy shell correction.
Evaluating their ratio within the extended Thomas-Fermi
effective-surface approximation, one finds 
good agreement with the quantum calculations.



\vspace{-0.2cm}
\section{INTRODUCTION}

Many theoretical approaches for nuclear rotations are 
based  on the Inglis cranking model  and  
Strutinsky shell-correction method (SCM) \cite{strutMOP}, 
extended to the rotational problems
by Pashkevich and Frauendorf \cite{pashfrau1975,mikhailovepnp}. 
For a deeper understanding of the correspondence
between classical and quantum physics of such rotations, 
it is worth to analyze 
the shell components of the moment of inertia (MI) within the  
periodic-orbit theory (POT)  
\cite{gutz,StrMag1976,KMS1979,FKMS1998,book,MYAB2011}.
In this context, one should refer to 
Ref.\ \cite{KMS1979} for the semiclassical 
description of the so called 
``classical rotation'' as an alignment of the particle angular momenta 
along the symmetry axis.
The semiclassical extended-Gutzwiller approach 
\cite{StrMag1976,KMS1979,MKS1978} also was
applied successfully to the description of 
the magnetic susceptibilities in 
metallic clusters  and quantum dots as a Landau diamagnetic response
\cite{FKMS1998} (see also 
Refs.\ \cite{Richter1996,MSKB2010,belyaev,GMBBPS2015}).
The perturbation expansion of Creagh  
\cite{book} has been used in the POT calculations of the MI shell 
corrections for
the spheroidal-cavity mean field \cite{DFP2004}. The 
semiclassical nature of the cranking model  imposes 
conditions of high angular momenta at larger nuclear deformations.
The 
non-perturbative Gutzwiller POT 
\cite{gutz,StrMag1976}, extended to the bifurcation phenomena 
at large deformations  \cite{MYAB2011,MAFM2002},
was  applied \cite{MSKB2010} 
to adiabatic (statistical-equilibrium) collective rotations around an
axis perpendicular to the symmetry axis in the case of 
the harmonic-oscillator mean field. The MI for such rotations is 
described as the sum of the 
Extended Thomas-Fermi (ETF) MI $\Theta _{\rm ETF} $  
\cite{bartel,belyaev} and shell corrections 
$\delta \Theta $  \cite{MSKB2010,belyaev}. 
By including self-consistency and spin effects into the MI 
calculations, a more realistic description of collective rotations 
is obtained within the  ETF approach \cite{bartel,ETF_BGH1985}.
A phase-space trace formula for the MI shell components $\delta \Theta$ 
was obtained \cite{GMBBPS2015} in 
terms of the free-energy shell corrections $\delta F$, for 
integrable highly idealized Hamiltonians such as the deformed 
harmonic oscillator \cite{MSKB2010} and the spheroidal cavity
\cite{MYAB2011,MAFM2002,GMBBPS2015}. Spin and pairing effects, 
as well as higher order $\hbar^2$ corrections were however neglected  
\cite{GMBBPS2015}. In the present work,  
$\hbar^2$ surface corrections to the ratio $\delta \Theta/\delta F$
are taken into account   
within the ETF model in the (leptodermous) effective surface (ES)
approach \cite{strmagden,BMRV,BMR_PS2015,BMR_PRC2015}. 

\vspace{-0.2cm}
\section{CRANKING MODEL AND SHELL-STRUCTURE}
 
Within the cranking model, the nuclear collective rotation 
of an independent-particle Fermi system 
is associated with an 
eigenvalue problem for
the many-body Hamiltonian (Routhian), 
  $\widehat{H}^{}_{\om} = \widehat{H} \!-\! \om \widehat{\ell}_x$,  
where $\hat{\ell}_x$ is the operator for the 
particle angular-momentum projection onto 
the $x$ axis, perpendicular to the symmetry $z$ axis.
The frequency $\om$ and the chemical potential 
$\lambda$, 
which are the Lagrange multipliers of the constrained variational problem,
are determined by the
angular momentum projection $I_x$ onto this $x$ axis 
and the particle number conservation $N$.
The MI $\Theta_x$ can 
be considered as a susceptibility 
(Refs.\ \cite{FKMS1998,Richter1996,MSKB2010,belyaev,GMBBPS2015}):
\be\l{MIdef} 
\Theta_{x}=
\partial \langle \hat{\ell}_x\rangle_\om/\partial \om
=\partial^2 E(\om)/\partial \om^2\;,
\ee
where $E(\om)$ is the quantum average of the Hamiltonian $\widehat{H}$, 
i.e.\ the
energy of the yrast line $E(I_x,N)$ resulting from these 
two constraints. Using the coordinate representation 
for the MI $\Theta_x$ in terms of the 
one-body semiclassical Gutzwiller expansion for the
Green's function, 
for the adiabatic statistically equilibrium 
rotations in the nearly local approximation one 
obtains the MI 
phase-space trace formula \cite{GMBBPS2015}:
\bea\l{misclps}
 &&\Theta_{\rm scl} \approx d_s \, m \int {\rm d} \eps\; \eps\; n(\eps)
\int \frac{{\rm d} \r {\rm d} \p}{(2 \pi \hbar)^3}
\frac{r_{\perp}^2}{\eps} \nonumber\\
&\times& g_{\rm scl}(\r,\p;\eps) 
= \Theta_{\rm ETF} +\delta \Theta_{\rm scl}\;,
\eea
where $m$ is the nucleon mass, $n(\eps)$ the occupation number, $d_s$
the spin (spin-isospin) degeneracy, and $r_{\perp}^2\!=\!y^2+z^2$ in
Cartesian coordinates.
Starting from the Wigner distribution function $f(\r,\p)$, one defines 
the one-body density $g(\r,\p;\varepsilon)$ in 
the phase 
space $\r,\p$ and 
energy $\eps$ as the derivative of $f(\r,\p)$ with
    respect to $\varepsilon$ [see Eq.\ (\ref{laplac})]. This density
$g$ can be
   written, like traditionally done, as \cite{MSKB2010,GMBBPS2015,belyaev}
\be\l{gscl}
g_{\rm scl}(\r,\p;\eps)=
g^{}_{\rm ETF}(\r,\p;\eps) + \delta g(\r,\p;\eps)\;,
\ee
where $g^{}_{\rm ETF}(\r,\p;\eps)$ is the ETF component and
    $\delta g(\r,\p;\eps)$ the shell correction 
(see Ref.\ \cite{GMBBPS2015}
    for 
the relation of 
$g_{\rm scl}(\r,\p;\eps)$ to the Gutzwiller Green's 
function expansion over classical trajectories).

In what follows we shall take advantage of the
strong resemblance of 
the MI (\ref{misclps})  
with the  
semiclassical single-particle energy.  
 The only difference is that an 
additional factor 
$~m r_{\perp}^2/\eps~$
appears in Eq.\ (\ref{misclps}).
The same subdivision in terms of the 
ETF and shell 
components is obtained at finite temperatures $T$ 
after a statistical averaging in Eq.\ (\ref{misclps}) where
\bea\l{dmiF}
 && \delta \Theta_{\rm scl} \approx  
           m \, \langle r_{\perp} ^2 /\varepsilon \rangle\; 
\delta F_{\rm scl}\;,\nonumber\\
&& \delta F_{\rm scl} = \Re\sum_{\rm PO} 
\frac{\pi t_{\rm PO} T /\hbar}
{ \sinh (\pi t_{\rm PO} T / \hbar)}\;\delta E_{\rm PO}\;.
\eea
Brackets $\langle \cdots \rangle$ indicate an average over 
the variables $\r$, $\p$, and  
$\eps$ with a weight $\eps$, i.e.,
\be\l{psav}
\langle \, \frac{r_{\perp}^2}{\varepsilon} \,\rangle =
  \frac{\int {\rm d} \eps\; \eps\; n(\eps) 
\int {\rm d} \r \, {\rm d} \p \; \frac{r_{\perp}^2}{\varepsilon}
g_{\rm scl}(\r,\p;\eps)}{
\int {\rm d} \eps\; \eps\; n(\eps) 
\int {\rm d} \r \, {\rm d} \p\; g_{\rm scl}(\r,\p;\eps)} \;\; .
\ee
In Eq.\ (\ref{dmiF}), $\delta F_{\rm scl}$ 
is the semiclassical free-energy shell correction and 
$\delta E_{\rm PO}$  
the  periodic-orbit (PO) component of the 
energy shell correction,
\bea\l{desclpo}
 && \delta E \approx \delta E_{\rm scl} = \Re\sum_{\rm PO} \delta E_{\rm PO}
  \quad  \mbox{with}  \quad\nonumber\\ 
&&\delta E_{\rm PO} \propto \frac{\hbar ^2}{t_{\rm PO}^2} \;
\exp\left[\frac{i}{\hbar}\; S_{\rm PO}(\lambda) - \frac{i\pi}{2}\;
\mu^{}_{\rm PO}\right]\;.
\eea
The period $~t^{}_{\rm PO}~$,  and  
the action $~S_{\rm PO}(\eps)~$ 
for the particle motion along the PO are taken 
at the chemical potential $\eps\!=\!\lambda \!\approx\! \eps^{}_F$ 
(at $\om\!=\!0$ and  
$T\!=\!0$) where $\eps^{}_F$  is the Fermi energy  
\cite{book,MYAB2011,belyaev}. The Maslov phase 
$\mu^{}_{\rm PO}$ is determined
by the number of the caustic and turning points along the PO. POs appear 
by the improved stationary phase method (ISPM) through
integrations over the phase space variables 
\cite{MYAB2011,MAFM2002,belyaev,GMBBPS2015}. For the phase-space average 
$\langle r_{\perp}^2 / \eps \rangle$ in Eq.\ (\ref{psav}) 
one again obtains approximately 
a decomposition into ETF and 
shell-correction contributions through the distribution function
$g_{\rm scl}(\r,\p;\eps)$.

\vspace{-0.2cm}
\section{SURFACE CORRECTIONS}

Using the inverse 
Laplace transformation (\ref{laplac}) 
one arrives at an expansion 
up to order $\hbar^2$ 
of the smooth semiclassical one-body 
distribution function (Appendix A), 
\be\l{getf}
g^{}_{\rm ETF}(\r,\p;\eps)\approx
g^{}_{\rm TF}(\r,\p;\eps) + g^{}_{\rm S}(\r,\p;\eps)\;, 
\ee
with
the TF and surface components, 
\be\l{ftf}
g^{}_{\rm TF}(\r,\p;\eps) = \delta \left( \eps - H_{\rm cl}(\r,\p) \right)\;,
\ee
%
\bea\l{fetf}
&&g^{}_{\rm S}(\r,\p;\eps) =\hbar^2
\left\{-\frac{\nabla^2V}{4m}\;\frac{
\partial^2\delta\left(\eps-H_{\rm cl}(\r,\p)\right)}{\partial \eps^2}\right.
\nonumber\\ 
&+&\left. \left[\frac{\left(\nabla V\right)^2}{6m}
+ \frac{\left(\p \nabla\right)^2V}{6 m^2}\right]\;\frac{
\partial^3\delta\left(\eps-H_{\rm cl}(\r,\p)\right)}{\partial \eps^3}\right.
\nonumber\\
&-&\left.\frac{\left(\p \nabla V\right)^2}{8m^2}\;
\frac{
\partial^4\delta\left(\eps-H_{\rm cl}(\r,\p)\right)}{\partial \eps^4}
\right\}\;.
\eea
 \\[ -2.0ex]
Here $H_{\rm cl}(\r,\p)\!=\!p^2/(2m)\!+\!V(\r)$ is the 
classical Hamiltonian with the mean-field potential $V(\r)$ .
Gradients of the potential V in the surface correction $g^{}_{\rm S}$
     of order $\hbar^2$ can be expressed, within the ETF method 
\cite{ETF_BGH1985,book},
     to the same $\hbar^2$ order, in terms of gradients
of the TF particle density [see Eqs.\ (\ref{dv2}) and (\ref{d2v})],
\be\l{rhoTF} 
\rho^{}_{\rm TF}=d_s [2m (\lambda-V(\r))]^{3/2}/(6 \pi^2 \hbar^3)\;.
\ee

 From Eqs.\  (\ref{gscl}), (\ref{getf})--(\ref{fetf}) 
and (\ref{psav}) one obtains, for the spheroidal cavity, within the 
 ETF  ES approximation up to $\hbar^2$ corrections, 
\be\l{ell2ms}
  \langle r_{\perp}^2 / \eps \rangle^{}_{\rm ETF} \approx 
\frac{a^2+b^2}{3 \lambda}\; 
\frac{1+\Theta_{\rm S}/\Theta_{\rm TF}}{1+E_{\rm S}/E_{\rm TF}} \;,
\ee
where $a$ and $b$ are the semi-axes of the spheroid.
Imposing volume conservation requires that $a^2 b=R^3$,
where $R$ is the radius of  the equivalent sphere.
$E_{\rm TF}$, $\Theta_{\rm TF}$ and $E_{\rm S}$, $\Theta_{\rm S}$ 
are the TF and $\hbar^2$ ETF surface 
components, 
respectively \cite{book}.
The surface energy $~E_{\rm S}\!=\!\sigma \mathcal{S}\!=\!b_{\rm S}N^{2/3}~$ 
with the  
spheroid area $\mathcal{S}$ and  
surface energy constant $b_{\rm S}=4 \pi r_0^2 \sigma$ is determined by the
 surface tension $\sigma$ of the capillary pressure. Within the  ETF  model,  
$\sigma$ is defined by the 
$\hbar^2$ correction to the kinetic energy (Appendix B),  
\be\l{sigma0}  
\sigma=\frac{\hbar^2}{72 m}\; \int_{-\infty}^{\infty} 
\frac{\d \xi}{\rho^{}_{\rm TF}}\;
\left(\frac{\partial \rho^{}_{\rm TF}}{\partial \xi} \right)^2\;,
\ee
 where $\xi$ is locally the distance from a given point $\r$ to the ES 
\cite{book,belyaev,strmagden}.  
The surface corrections in Eq.\ (\ref{ell2ms}) are given by
\bea\l{Estf}
  {}\hspace{-0.2cm}
 \frac{E_{\rm S}}{E_{\rm TF}} &=&
  \frac{5b^{}_{\rm S}\mathcal{S}}{12\pi \eta^{2/3} a^2\lambda\;N^{1/3}} \qquad
\mbox{and} \nonumber\\ 
\frac{\Theta_{\rm S}}{\Theta_{\rm TF}} &=&
  5b^{}_{\rm S}\; \frac{\eta^2I_0+\pi(1-2\eta^2)I_1}{
  \pi^2 \eta^{2/3}(1+\eta^2)\lambda\;N^{1/3}}\;,
\eea
where
$$
I_0=1+
\frac{\eta^2 \arctan\sqrt{\eta^2-1}}{\sqrt{\eta^2-1}}
$$
                                                               \\[ -3.0ex]
and 
\bea\l{I1}
  I_1 &=& \frac{2 \eta}{3 \sqrt{\eta^2-1}} \left[(2\eta^2-1)
E\left(\sqrt{\eta^2-1}/\eta\right)\right.\nonumber\\
&-& \left.
K\left(\sqrt{\eta^2-1}/\eta\right)\right]
\eea
                                                                   \\[ -3.0ex]
with
$E(\kappa)$ and $K(\kappa)$ 
being 
the complete elliptic integrals \cite{byrd}.  
The deformation parameter is given by $\eta=b/a$. 
In units of the classical rigid-body (TF) MI, 
$\Theta_{\rm TF}\!=\!m \left(a^2+b^2\right) N/5$,
one finally obtains 

\vspace{0.2cm}
\be\l{dtxtxtf}
  \frac{\delta \Theta_x}{\Theta_{\rm TF}} = \frac{5 \left(1 
        +\Theta_{\rm S}/\Theta_{\rm TF}\right)}{1+ E_{\rm S}/E_{\rm TF}}  
                      \frac{\delta F}{3 N \lambda} \;.
\ee
%

\vspace{-0.2cm}
\section{DISCUSSION OF RESULTS}

\begin{figure*}
\begin{center}
\includegraphics[width=0.7\textwidth]{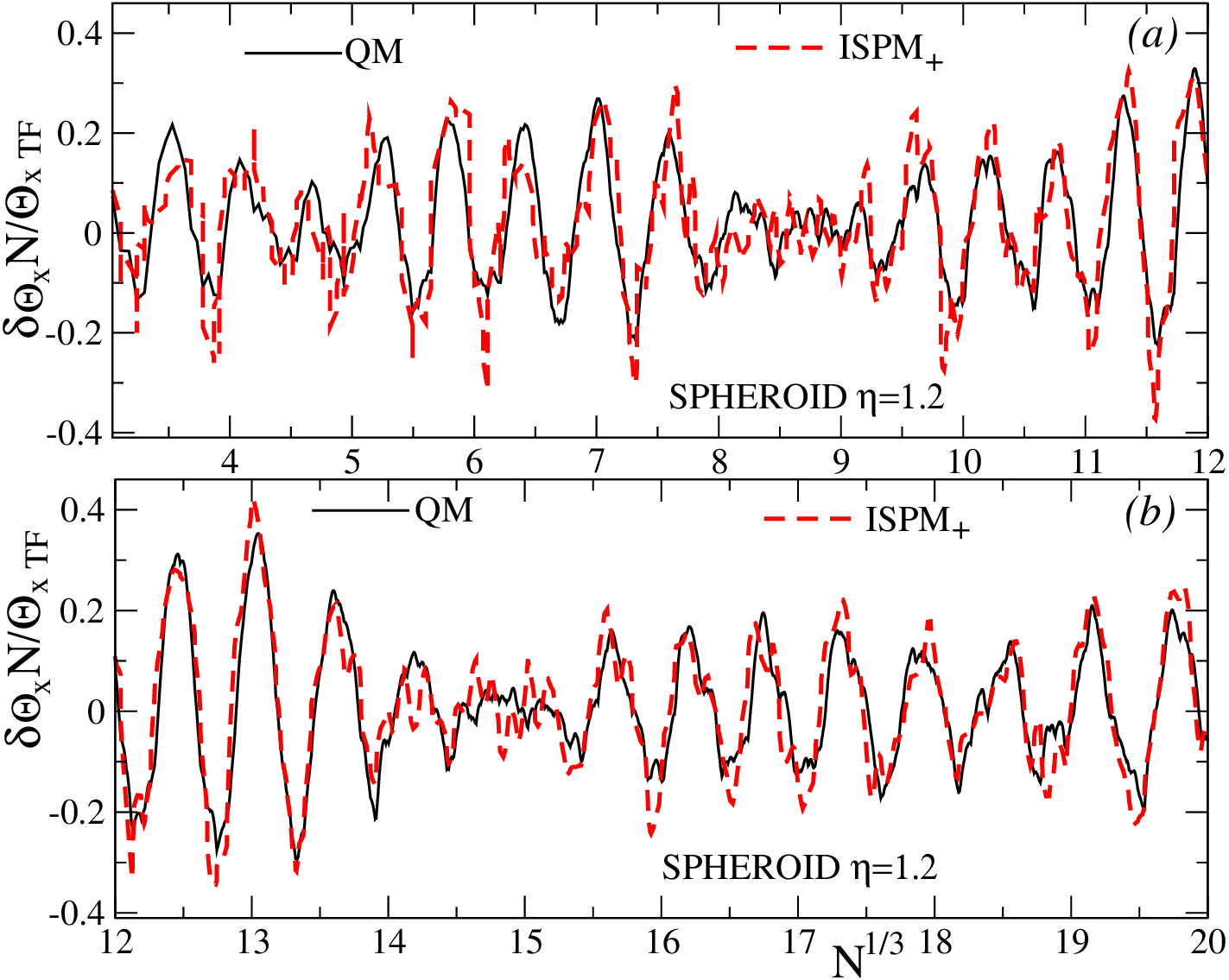}
\end{center}

\vspace{-0.2cm}
\caption{
{\it (Color online)} MI shell components $\delta \Theta_x$ 
(in TF units) as function of
$N^{1/3}$ at deformation $\eta=b/a=1.2$ obtained in a quantum-mechanical
  (QM) and a semiclassical calculation, including surface corrections
 (ISPM$_+$) for smaller (upper part (a)) and larger (lower part (b))
  particle numbers.
}
\label{fig1}
\end{figure*}

\begin{figure*}
\begin{center}
\includegraphics[width=0.7\textwidth]{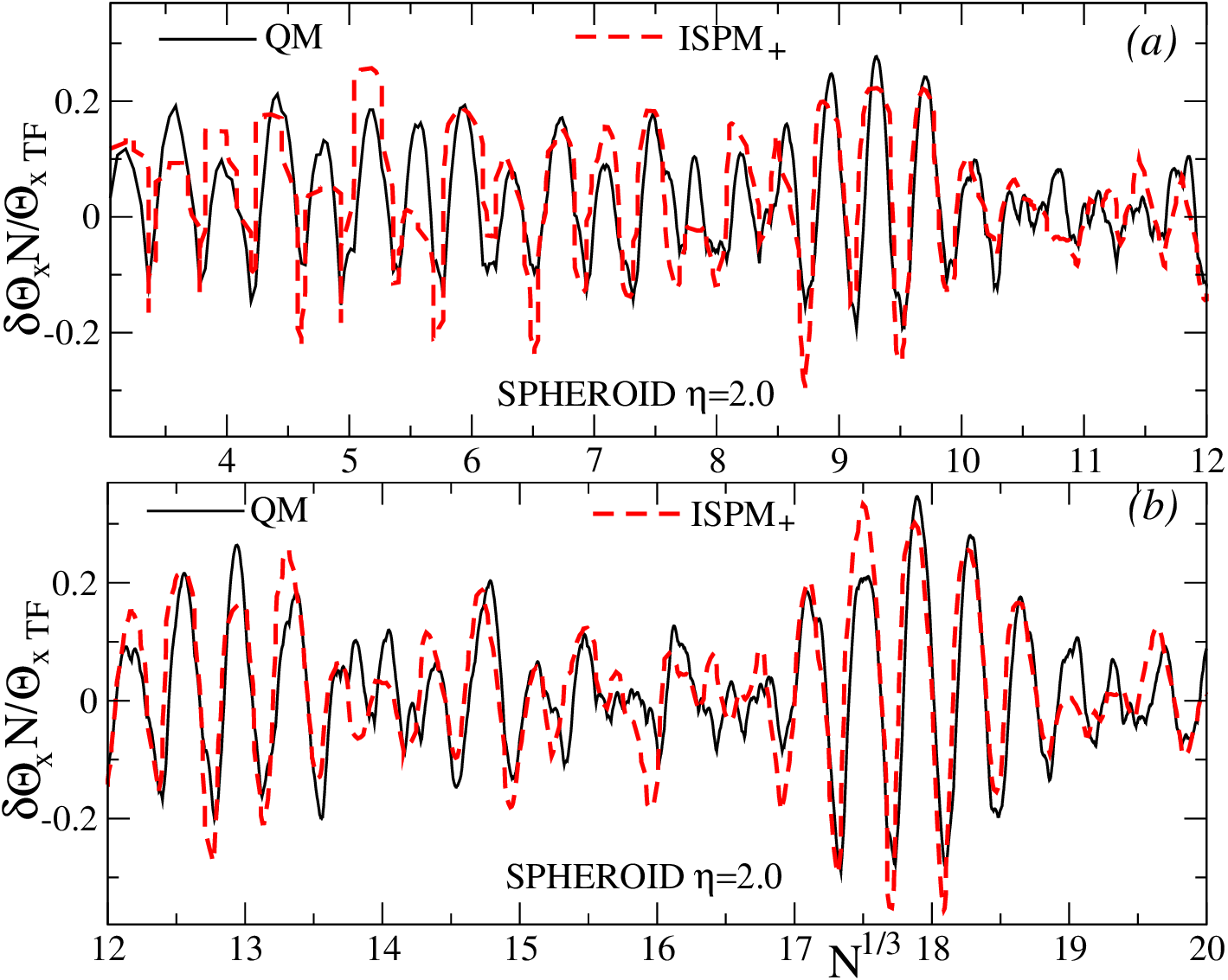}
\end{center}

\vspace{0.2cm}
\caption{{\it (Color online)} Same as Fig.\ \ref{fig1} but for a 
deformation of $\eta =2.0$.
} 
\label{fig2}
\end{figure*}

\begin{figure*}
\begin{center}
\includegraphics[width=0.7\textwidth]{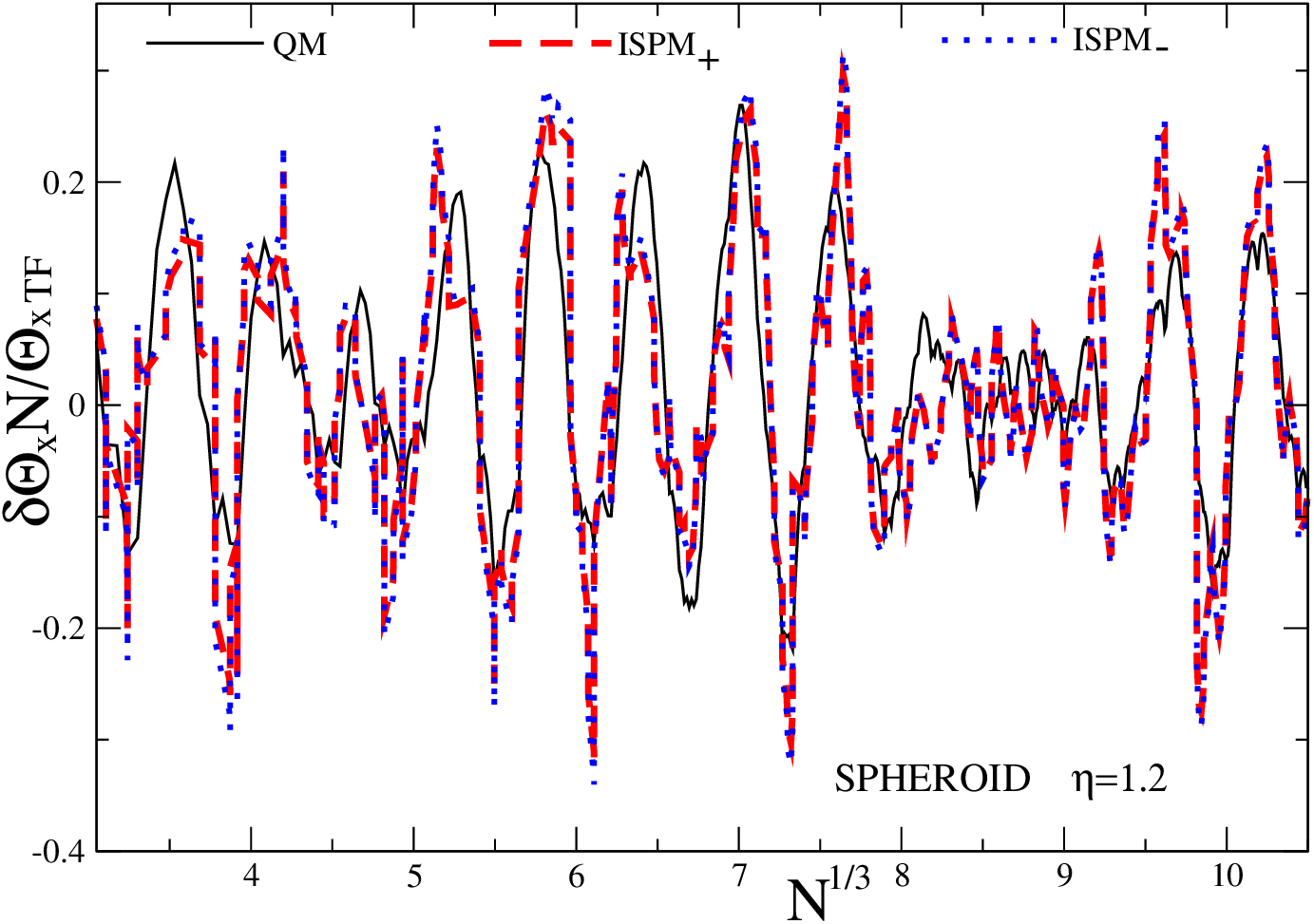}
\end{center}

\vspace{0.1cm}
\caption{
{\it (Color online)}  Comparison between the MI shell 
components $\delta \Theta_x$ (in TF
units) obtained with (ISPM$_+$) and without (ISPM$_-$) surface
   corrections as function of $N^{1/3}$. For comparison the quantum result
(black solid line) is also shown. The deformation is $\eta = 1.2$.
} 
\label{fig3}
\end{figure*}

\begin{figure*}
\begin{center}
\includegraphics[width=0.7\textwidth]{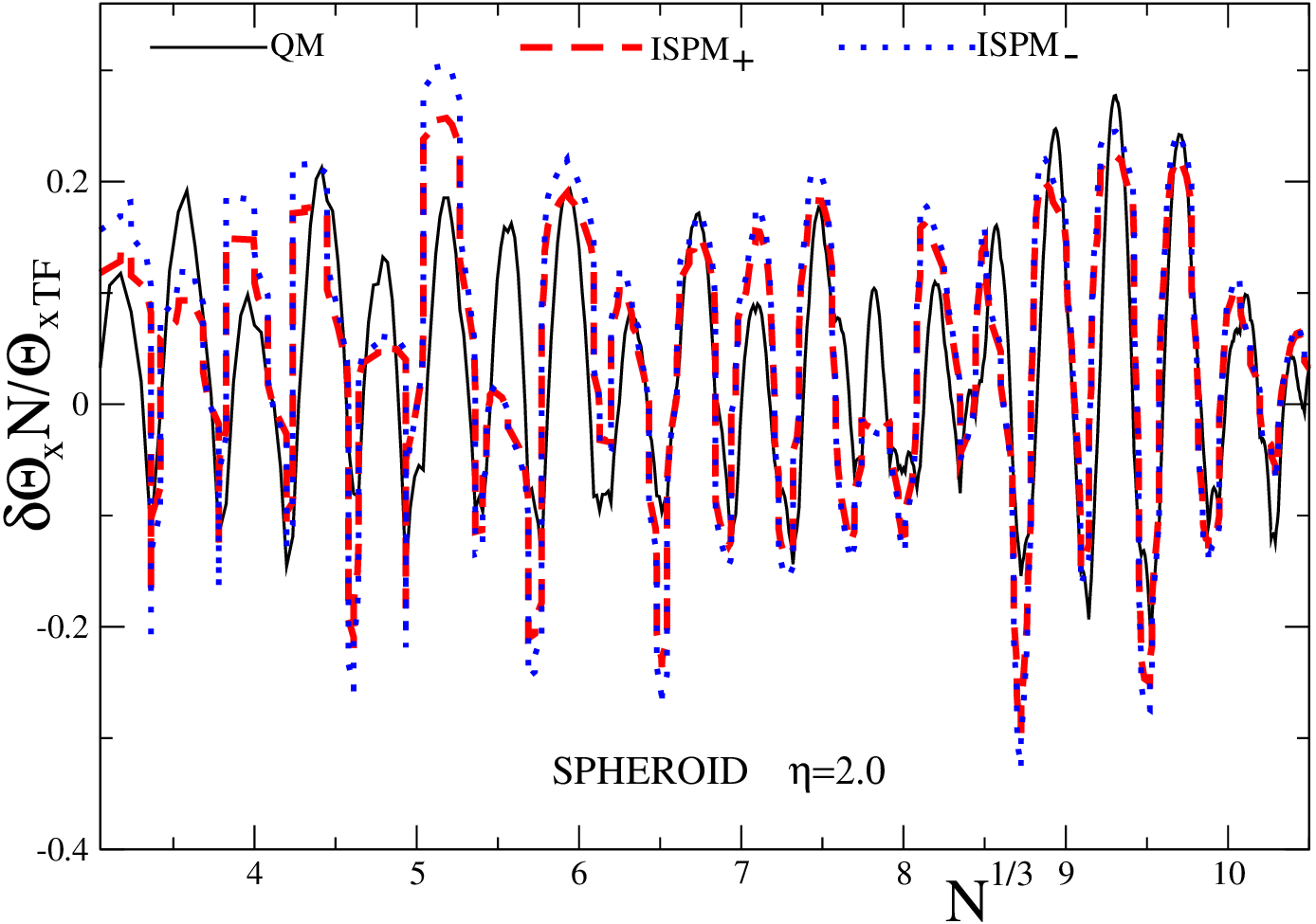}
\end{center}

\vspace{0.1cm}
\caption{
{\it (Color online)}  Same as Fig.\ \ref{fig3} 
but for $\eta =2.0$.
} 
\label{fig4}
\end{figure*}

Figures \ref{fig1} and \ref{fig2} show a comparison between the 
semiclassical ISPM 
MI shell corrections (\ref{dmiF}) obtained 
with (index $+$)  
surface terms and the
quantum-mechanical (QM) result.
The latter is determined 
through the ETF average (\ref{ell2ms}) for 
$\langle r_{\perp}^2/\eps \rangle$ with a realistic surface energy constant
$b^{}_{\rm S}\approx 20 $~MeV 
whereas the energy shell correction $\delta E$   
(equal $\delta F$ at zero
 temperature $T$) is calculated 
by the  SCM 
using the quantum spectrum. A large supershell
effect appears in 
$\delta \Theta_x$, especially for larger 
deformations
 in the PO bifurcation region (Figs.\ \ref{fig2} and \ref{fig4}). 
The effect of the surface correction, Eq.\ (\ref{Estf}), 
is analyzed in
   Figs.\ \ref{fig3} and \ref{fig4} that show, together with the 
result of the quantum
  calculation, the shell components $\delta \Theta_x/\Theta_{\rm TF}$
   obtained with (ISPM$_{+}$) and without (ISPM$_{-}$) these surface
   corrections. The difference between both curves is seen to be more
   important for small particle numbers, which can be easily understood
   since the surface corrections decreases as $N^{-1/3}$ as seen from Eq.\
   (\ref{Estf}).
The contribution of the shorter three-dimensional orbits bifurcated from the 
equatorial ones are dominating in the case of large deformations 
(Fig.\ \ref{fig2}), in contrast to the small deformation region where
the meridian orbits are predominant (Figs.\ \ref{fig1} and \ref{fig3}), 
in accordance with
Refs.\ \cite{MYAB2011,MAFM2002}.
One also observes that
the surface corrections become more significant  
with increasing
deformation of the system.

 For small temperatures one has 
$\delta F_{\rm scl}\!\approx\!\delta E_{\rm scl}$, 
and therefore, a remarkable  interference of the 
dominant short three-dimensional and meridian 
 orbits is  shown in Refs.\ 
\cite{MYAB2011,MAFM2002,GMBBPS2015}.  
Their bifurcations in the superdeformed region
give essential contributions to the MI through the
      (free) energy shell corrections.
With increasing temperature the shorter 
equatorial orbits become dominating, 
as seen analytically from the exponentially decreasing 
temperature-dependent factor in Eq.\ (\ref{dmiF}).

The shell corrections (\ref{dmiF}) to the MI  are  
relatively much smaller than the classical rigid-body (TF) 
component. This is similar to the (free) energy shell corrections $\delta E$  
(or $\delta F$) as compared with the ETF volume and surface energy. 
However, many 
important physical effects, such as fission isomerism
and high spin physics depends basically  
on the shell effects.
Our non-perturbation results for the MI shell corrections can  
be applied for larger rotational frequencies and larger deformations 
$\eta\!\sim\!1.5\!-\!2.0$ where the 
bifurcations play the dominating role.

\vspace{-0.2cm}
\section{SUMMARY AND CONCLUSIONS}

Within the non-perturbative Gutzwiller POT we derived the MI 
shell component $\delta \Theta$ in terms of the free-energy shell correction 
$\delta F$ for any mean-field potential by taking into account the ETF 
$\hbar^2$ corrections in the effective surface approximation.
For the deformed spheroidal cavity,  we found a good agreement
between the semiclassical POT and quantum results for 
$\delta \Theta$ at several  
deformations and temperatures. The surface corrections 
become more significant with increasing deformations and decreasing 
particle numbers. With increasing temperature, one finds the
 generally observed exponential
decrease of the shell effects. 
For large deformations and small temperatures, 
one observes some remarkable supershell effects due to  
the interference of three-dimensional and meridian orbits bifurcating
from the equatorial orbits.

For future research in this field,
it would be valuable to include the neutron-proton asymmetry 
\cite{BMRV,BMR_PS2015,BMR_PRC2015} and the spin degrees of freedom into the 
semiclassical MI shell calculations \cite{belyaev}. The latter lead 
to the well-known spin-orbit
splitting which significantly changes the nuclear shell structure  
and accounts for spin paramagnetic effects \cite{belyaev}. 
The MI expressions obtained analytically at the present stage have therefore 
only a somewhat restricted  values 
for the use in real nuclei, but could be 
directly applied to the magnetic susceptibility 
for metallic clusters and quantum dots \cite{FKMS1998}. 
The extension of the POT  to the MI 
shell correction calculations with 
the inclusion of 
the spin degree of freedom would  
constitute an essential progress in 
understanding  
the relation between the nuclear MI and the free-energy shell 
corrections. For a more realistic study, let 
us also mention the inclusion of pairing correlations, especially far from 
deformed magic nuclei and non-adiabatic effects.
 The work along these lines is in progress.

\vspace{-0.2cm}
\bs 
\centerline{{\bf ACKNOWLEDGMENTS}}
\ms 
We would like to thank K.\ Arita, M.\ Brack, M.\ Matsuo, 
K.\ Matsuyanagi, 
and K.\ Pomorski for helpful and 
stimulating discussions. One of us (A.G.M.) is
also very grateful for nice hospitality and financial support 
during his working visits of the
National Centre for Nuclear Research (Poland),
the Strasbourg Institut Pluridisplinaire Hubert Curien (France), the Nagoya
Institute of Technology (Japan), and
Japanese Society of Promotion of Sciences, 
Grant No.\ S-14130. 
%

\appendix \setcounter{equation}{0}
\renewcommand{\theequation}{A\arabic{equation}}

\vspace{-0.2cm}
\begin{center}
\textbf{Appendix A: THE WIGNER-KIRKWOOD METHOD
}
\label{appA}
\end{center}

The Wigner-Kirkwood method starts with the Gibbs operator \cite{book},
$\hat{C}_\beta\!=\!\exp(-\beta \hat{H})$,
where $\hat{H}$ is the quantum-mechanical Hamiltonian.  
In the case that
$\hat{H}$ is time independent, the coordinate-space  
representation of the Gibbs operator, the so-called Bloch density matrix,
  is given by
\be\l{gibbscoor}
C(\r^{}_1,\r^{}_2;\beta)=
\sum_i\psi^\ast_i(\r^{}_1)\;\exp(-\beta\;\varepsilon_i)\;\psi_i(\r^{}_2),
\ee
where $\psi_i$ and $\varepsilon_i$ are the eigenfunctions and eigenvalues
of the Hamiltonian ($\hat{H} \psi_i\!=\!\varepsilon_i \psi_i$). 
Therefore, after  formally replacing $\beta\!=\!it/\hbar$, the  
Bloch density matrix $C(\r^{}_1,\r^{}_2;\beta)$
is seen to be nothing but the  one-body time-dependent 
propagator (Green's function)  
$K(\r^{}_1,\r^{}_2;t)$ and one can use
the corresponding Schr{\"o}dinger
equation for the calculation of $C(\r^{}_1,\r^{}_2;\beta)$ \cite{book}.
Note that the POT in the extended Gutzwiller version 
starts with the solution of this equation for the
propagator $K(\r^{}_1,\r^{}_2;t)$ in terms of the 
Feynman path integral. Its calculation by the stationary phase method 
leads to the semiclassical expression for $K(\r^{}_1,\r^{}_2;t)$, 
and then, one can get
the semiclassical expansion of the Green's function, 
$G(\r^{}_1,\r^{}_2; \varepsilon)$,
and its traces,  
namely the level density, $g(\varepsilon)$, and the 
particle density $\rho(\r)$ 
(at $\r^{}_1\!\rightarrow\! \r^{}_2\!=\!\r$).
The shell components of these densities can be
 expressed in terms of the closed trajectories (see the main text
for the case of the oscillating level-density part written in terms
of POs). 
Thus, the POT can be developed for the Bloch density matrix 
$C(\r^{}_1,\r^{}_2;\beta)$ itself.

In order to solve semiclassically the Schr{\"o}dinger equation for 
the Bloch function
$C(\r^{}_1,\r^{}_2;\beta)$, one can make a transformation, first from 
$\r^{}_1$ and $\r^{}_2$ to the center-of-mass and relative coordinates, 
$\r\!=\!(\r^{}_1\!+\!\r^{}_2)/2$ 
and $\s\!=\!\r^{}_2\!-\!\r^{}_1$, and then, by the Fourier transformation to the
phase-space variables, $\{\r,\p\}$,  
what corresponds to a Wigner 
transformation from 
$C(\r^{}_1,\r^{}_2;\beta)$ to $C_W(\r,\p;\beta)$,
\bea\l{wigner}
C_W(\r,\p;\beta)&=&
\int \frac{\d \s}{(2 \pi \hbar)^3}\; C(\r-\s/2,\r+\s/2;\beta)\nonumber\\
&\times&\exp\left(i\p \s/\hbar\right)\;.
\eea
This reduces one complicated Schr\"odinger equation
to an infinite system of much simpler first-order ordinary
differential equations
(at each power of $\hbar$, see Ref.\ \cite{book}) 
which can be analytically integrated.

The advantage of the Wigner-Kirkwood method is obviously to generate
   smooth quantities averaged over many quantum states to smooth out quantum
   oscillations like shell effects.
The POT on the contrary is aimed at 
the derivation
of analytical expressions for the shell components of the partition 
function, and thereby of the level and  particle densities. 
In the Wigner-Kirkwood method, the main term of the expansion
of $C_W(\r,\p;\beta)$ is proportional to the classical distribution function 
$f_{\rm cl}(\r,\p)$, and $\hbar$ corrections can be obtained 
by solving
a simple system of differential equations at each power of $\hbar$.
 Strictly speaking there is no convergence of this asymptotical 
expansion because
of presence of the $\hbar$ in the rapidly oscillating exponents.  
Therefore, to get the convergent series in $\hbar$
of the ETF approach, 
one first has to use local averaging in the phase 
space variables and then, 
expand smooth quantities in a $\hbar$ series,  
in contrast to the shell-structure
POT. In this way, the simple 
ETF $\hbar$ expansions of    local quantities such as the
   particle density   $\rho(\r)$, kinetic energy density   $\tau(\r)$, and
   level density   $g(\varepsilon)$ are obtained.

The canonic partition function $\mathcal{Z}(\beta)$  is
 obtained by integrating over the whole space the diagonal Bloch matrix 
$C(\r,\r;\beta)\!=\!C(\r;\beta)$,
\be\l{partfun}
\mathcal{Z}(\beta)=\int \d \r\; C(\r;\beta)=
\sum_i \exp (-\beta \varepsilon_i)\;.
\ee
The trace, $\mathcal{Z}\!=\!\mbox{Tr} \{\exp(-\beta \hat{H})\}$, 
can be taken for any complete set of states. 
For the semiclassical expansion involving an integral 
over the phase
space, it is more convenient
to take plane waves as the complete set. We may then write
\be\l{partfunps}
\mathcal{Z}(\beta)=
\int \frac{\d \r\; \d \p}{(2 \pi \hbar)^3} e^{-i \p \r/\hbar}\;
e^{-\beta \hat{H}}\;e^{i \p \r/\hbar}\;.
\ee
As the kinetic operator in $\hat{H}$ does not commute with the potential
$V(\r)$, it is convenient to use the following representation \cite{book}:
\be\l{wudef}
e^{-\beta \hat{H}}e^{i \p \r/\hbar}=e^{-\beta H_{\rm cl}}\;
e^{i \p \r/\hbar}\;w(\r,\p;\beta)\;,
\ee
where $H_{\rm cl}$ is the classical Hamiltonian that appears in
 Eqs.\ (\ref{ftf}) and (\ref{fetf}). Solving
the Schr{\"o}dinger equation for the function $w$
with the boundary condition $\mbox{lim}^{}_{\beta \to 0}w(\r,\p;\beta)\!=\!1$,
one assumes that $w(\r,\p;\beta)$ can be expanded in  
a power series in 
$\hbar$:
\be\l{wexp}
w=1+\hbar w_1 + \hbar^2 w_2 + \cdots \;.
\ee
Equating
terms of the same power in $\hbar$ from both sides of this differential
equation, one obtains 
the $\hbar$ corrections:
\begin{equation}\label{w1}
w_1=-\frac{i\beta^2}{2m}\;\p\cdot \nabla V, \hspace{1.8cm}  
\end{equation}
and
\bea\label{w2}
&&w_2=-\frac{\beta^2}{4m} \nabla^2V+ \frac{\beta^3}{6m} 
\left(\nabla V\right)^2 \nonumber\\
&-&\frac{\beta^4}{8m^2} \left(\p\cdot \nabla V\right)^2
+\frac{\beta^3}{6m^2} \left(\p\cdot \nabla\right)^2 V\;.
\eea
The semiclassical series for the  partition function takes then the form:
\bea\l{partfunpsexp}
\mathcal{Z}(\beta)&=&
\int \frac{\d \r\; \d \p}{(2 \pi \hbar)^3} e^{-i \p \r/\hbar}\;
e^{-\beta H_{\rm cl}}\nonumber\\
&\times&\left(1+\hbar w_1 + \hbar^2 w_2 + ...\right)\;.
\eea
Differentiating the TF particle density $\rho^{}_{\rm TF}$ (\ref{rhoTF})
and solving the obtained linear
system of equations for the gradients 
of the potential, 
one finds
\be\l{dv2}
\left(\nabla V\right)^2= 
\left(\frac{\pi^2\hbar^2}{m (3\pi^2 \rho)^{1/3}}\right)^2
\left(\nabla \rho\right)^2,
\ee
\be\l{d2v}
\nabla^2V=\frac{\pi^2\hbar^2}{m (3\pi^2\rho)^{1/3}}
\left[
\frac{\left(\nabla \rho\right)^2}{3\rho}
-\nabla^2 \rho\right]
\ee
where the subscript TF on the density has been omitted. 
These expressions are more convenient to use in the more general case,
   including billiard systems, in particular, the spheroidal cavity.

For calculations of the semiclassical distribution function 
$g(\r,\p;\varepsilon)$, one can apply the inverse 
Laplace transformation:
\bea\l{laplac}
&& g(\r,\p;\varepsilon) = \frac{\partial f(\r,\p)}{\partial \varepsilon}
    = \frac{1}{2 \pi i}\; \int_{\beta_r - i \infty}^{\beta_r + i \infty} d\beta \; 
\nonumber\\
&\times&  \exp\left[\beta \left(\varepsilon - H_{\rm cl}\right)
\right]
 \left(1+\hbar w_1+\hbar^2 w_2\right), 
\eea
 where $w_1$ and $w_2$ are the semiclassical corrections 
of Eqs.\ (\ref{w1}) and (\ref{w2}).
 The integration in the complex $\beta$ plane in Eq.\ (\ref{laplac}) has to be
 taken along the imaginary axis, at a distance $\beta_r$ such that all
 singularities are located at its left. 
The linear term in $\hbar$, i.e. the term $w_1$ that is linear in $\p$,
does not
 contribute to the phase-space (momentum) integral for the energy $E$ and
 for the MI $\Theta$ in Eq.\ (\ref{misclps}). 
 Calculating the integral in Eq.\ (\ref{laplac}) using Eq.\ (\ref{w2}),
 one arrives, after some simple algebraic transformations, at Eq.\
 (\ref{fetf}).
\appendix \setcounter{equation}{0}
\renewcommand{\theequation}{B\arabic{equation}}

\vspace{-0.2cm}
\begin{center}
\textbf{Appendix B:  THE ES METHOD}
\label{appB}
\end{center}

For independent nucleons bound in a potential well, the  
energy density
 $\mathcal{E}(\rho)$ of symmetric nuclear matter 
($N\!=\!Z \!=\!A/2$) is found
to be \cite{strmagden,BMRV,BMR_PS2015,BMR_PRC2015}
\be\l{enerdenins} 
\mathcal{E}\left(\rho\right)=-b^{}_{\rm V}\rho + \rho\varepsilon(\rho) +
\Gamma\;\left(\nabla \rho\right)^2/(4 \rho)\;,
\ee
where $b^{}_V$ is the separation energy per particle, 
$\varepsilon(\rho)\!\approx\![K/18 \rho^2_{\infty}] 
\left(\rho\!-\!\rho_{\infty}\right)^2$, and 
where 
$K$ and $\rho_\infty $ are the incompressibility modulus and the
particle density of infinite nuclear matter, and 
$\Gamma\!=\!\hbar^2/(18m)$. 
For simplicity we neglect spin-orbit and asymmetry terms. A variation of 
the energy functional, $E\!=\!\int \d \r \;\mathcal{E}[\rho(\r)]$
with the energy density (\ref{enerdenins}) leads to the Lagrange equation
   \cite{strmagden}:
\be\l{lagreqs}
\frac{\Gamma}{2\rho}\; \triangle \rho -
\frac{\Gamma}{4 \rho^2} \left(\nabla \rho\right)^2 
-\frac{\d }{\d \rho}\left[\rho\;\varepsilon(\rho)\right]+\Lambda=0\;,
\ee
where $\Lambda=\lambda\!+\!b^{}_{\rm V}$ is the correction to 
the separation energy 
$-b^{}_{\rm V}$ in the chemical potential $\lambda$.  
This correction is 
proportional to a small leptodermous parameter $a/R\!\sim\!A^{-1/3} $
for heavy nuclei.
Introducing a local orthogonal-coordinate
system with a coordinate
   $\xi$ that defines the distance from a given point $\r$ to the effective
   surface (ES), one gets for the particle density $\rho_0$, in leading
   order in the leptodermous parameter $a/R$, a simple ordinary differential
   equation
\be\l{rhoeq0} 
\d \rho^{}_0/\d \xi =-2\rho^{}_0\; \varepsilon^{1/2}(\rho^{}_0)/\Gamma^{1/2}\;.
\ee
This equation can be solved analytically 
for the quadratic approximation to 
$\varepsilon(\rho)$.
Transforming the differential equation (\ref{rhoeq0}) to one for
   the dimensionless particle density, 
\bea\l{wxeps}
w(x)&=&\rho(\xi)/\rho_\infty\quad \mbox{with}\quad 
x = \xi/a\;,\nonumber\\
\epsilon(w) &=& (18/K)\; \varepsilon(\rho)=(1-w)^2\;,
\eea
   one finds

%
\be\l{weq0}
w'(x)=-\zeta w \sqrt{\epsilon(w)}\;, 
\ee
where $\zeta\!=\!2 a \sqrt{K/(18\Gamma)}$. 
Differentiating once more and using the fact that, by definition,
$w''(x)\!=\!0$ at the ES, one gets the boundary condition for
$w^{}_0\!=\!w(x=0)$:
\be\l{boundconds0}
2 \epsilon(w^{}_0) + w^{}_0\epsilon^\prime(w^{}_0)=0\;.
\ee
With Eq.\ (\ref{wxeps}) for $\epsilon(w)$, 
one finds the solution $w_0\!=\!1/2$.
Integrating Eq.\ (\ref{weq0}) 
using the boundary condition (\ref{boundconds0}), one obtains the 
explicit solution
\vspace{-0.2cm}
\be\l{wsols0}
w(x)=\left[1+\exp\left(\zeta x\right)\right]^{-1}\;,
\ee
 which tends asymptotically (for $x\!\rightarrow\!\infty$) to
  $w(x)\!\rightarrow\!\exp(-\zeta x)$. 
Therefore, one can define the diffuseness
parameter $a$ from the usual condition, 
$\zeta\!=\!1$ 
 so that the particle
   density $w(x)$ will be decreasing at large $x$ as $\exp(-x)$:
\be\l{difs0}
a=\sqrt{9 \Gamma/(2 K)}=\sqrt{\hbar^2/(4mK)}\;.
\ee
Another limiting case of finite constants of the 
potential part of the energy density 
[in front of $(\nabla \rho)^2$], including the spin-orbit and asymmetry
terms but neglecting
the kinetic energy term proportional to $\Gamma$
in Eq.\ (\ref{enerdenins}), was investigated in
Refs.\ \cite{BMRV,BMR_PS2015,BMR_PRC2015}.

For the energy $E$ 
with Eq.\ (\ref{enerdenins}), one has
\bea\l{energys}
E&=&-b_{\rm V}A + \int\d \r \left[
\frac{\Gamma}{4} \, \frac{\left(\nabla \rho\right)^2}{\rho}
+ \rho \varepsilon(\rho)\right]\nonumber\\
&=&E_{\rm V}+E_{\rm S}\;,
\eea
where $E_{\rm V}\!=\!-b_{\rm V}A$ is the volume and
$E_{\rm S}\!=\!\sigma S$ the surface component with the surface-tension
coefficient

\vspace{-0.5cm}
\be\l{sigmas}
\sigma= \frac{\Gamma}{2}\int_{-\infty}^{\infty} \frac{\d \xi}{\rho^{}_0}\;
\left(\frac{\partial \rho^{}_0}{\partial \xi}\right)^2\;.
\ee
For the calculation of the surface energy $E_{\rm S}$ from Eq.\
   (\ref{energys}), one needs the particle density $\rho\!\approx\!\rho_0$
   at leading order in the leptodermous parameter $a/R$. 
Due to the spatial derivatives in its integrand, and the definition of 
    $\epsilon(w)$ [Eq.\ (\ref{wxeps})], this surface integration gives, 
    in addition to the integrand, the contribution of order $a/R$.
   Therefore, according to the Lagrange equation at this order,
   Eq.\ (\ref{rhoeq0}), the two terms in square brackets in the integral
   in Eq.\ (\ref{energys}) turn out to be identical. Thus, we 
arrived at Eq.\ (\ref{sigmas}). Using Eqs.\ (\ref{rhoeq0})
   and (\ref{sigmas}) for the surface-tension coefficient, one finds
   (after transforming to dimensionless quantities and changing the
   integration variable from $x$ to $w$) the analytical result
\be\l{sigmas0}
\sigma=(\hbar \rho_{\infty}/36)\;\sqrt{K/m}\;.
\ee
Other limit cases are considered in Refs.\ \cite{BMRV,BMR_PS2015,BMR_PRC2015}.

\vspace{0.5cm}
%

\vspace{-2.0cm}

\vspace{1.0cm}


\vspace{-0.7cm}
\begin{thebibliography}{99}

\bibitem{strutMOP} V.M.\ Strutinsky {\it Nucl. Phys.} A {\bf 95} 420 (1967); 
1 (1968).

\bibitem{pashfrau1975} V.V.\ Pashkevich and S.\ Frauendorf,  
{\it Sov. J. Nucl. Phys.}  {\bf 20}, 588 (1975).

\bibitem{mikhailovepnp} I.N.\ Mikhailov, K.\ Neergard, V.V.\ Pashkevich, and 
S.\ Frauendorf, {\it Sov. J. Part. Nucl.} {\bf 8} 550 (1977).

\bibitem{gutz} 
M.\ Gutzwiller, J. Math. Phys. {\bf 12}, 343 (1971);
M.\ Gutzwiller, {\it Chaos in Classical and Quantum Mechanics} 
(Springer-Verlag, New York, 1990).

\bibitem{StrMag1976} V.M.\ Strutinsky, Nucleonika {\bf 20}, 679 (1975);
V.M.\ Strutinsky and A.G.\ Magner, Sov. J. Part. Nucl. {\bf 7}, 138 (1976).

\bibitem{KMS1979} V.M.\ Kolomietz, A.G.\ Magner, and V.M.\ Strutinsky,  
{\it Sov. J. Nucl. Phys.} {\bf 29} 758 (1979).

\bibitem{FKMS1998} S.\ Frauendorf, V.M.\ Kolomietz, A.G.\ Magner, and 
A.I.\ Sanzhur, {\it Phys. Rev.} B {\bf 58} 5622 (1998).

 \bibitem{book} M.\ Brack and R.K.\ Bhaduri, 
{\it Semiclassical Physics. Frontiers
in Physics}, No 96, 2nd ed. (Westview Press, Boulder, CO, 2003).


\bibitem{MYAB2011} A.G.\ Magner, Y.S.\ Yatsyshyn, K.\ Arita, and M.\ Brack, 
{\it Phys. At. Nucl.} {\bf 74} 1445 (2011).

\bibitem{MKS1978} A.G.\ Magner, V.M.\ Kolomietz, V.M.\ Strutinsky, 
Sov. J. Nucl. {\bf 28}, 764 (1978).

\bibitem{Richter1996} K.\ Richter, D.\ Ulmo, and R.A.\ Jalabert,
{\it Phys. Rep. } {\bf 276}, 1 (1996). 

\bibitem{MSKB2010} A.G.\ Magner, A.S.\ Sitdikov, A.A.\ Khamzin, and 
J.\ Bartel, {\it Phys. Rev.} C {\bf 81} 064302 (2010).

\bibitem{GMBBPS2015} D.V.\ Gorpinchenko, A.G.\ Magner, J.\ Bartel, 
and J.P.\ Blocki,
{\it Phys. Scr.}, T {\bf 90}, 114008 (2015).

\bibitem{belyaev} A.G.\ Magner, D.V.\ Gorpinchenko, and J.\ Bartel,
{\it Phys. At. Nucl.} {\bf 77} 1229 (2014).


\bibitem{DFP2004}M.A.\ Deleplanque, S.\ Frauendorf, V.V.\ Pashkevich et al.
{\it Phys. Rev.} {\bf C69} 044309 (2004).

\bibitem{MAFM2002} A.G.\ Magner, K.\ Arita, S.N.\ Fedotkin, 
and K.\ Matsuyanagi,
{\it Prog. Theor. Phys.} {\bf 108} 853 (2002).

\bibitem{bartel} K.\ Bencheikh, P.\ Quentin, and J.\ Bartel, Nucl. Phys. A 
{\bf 571}, 518 (1994).

\bibitem{ETF_BGH1985} M.\ Brack, C.\ Guet, and H.-B.\ H\aa kansson, 
{\it Phys.\ Rep.}\ {\bf 123}, 275 (1985).


\bibitem{strmagden} V.M.\ Strutinsky, A.G.\ Magner,
and V.Yu.\ Denisov, {\it Z. Phys.} A {\bf 322}, 149 (1985).

\bibitem{BMRV} J.P.\ Blocki, A.G.\ Magner, P.\ Ring, and A.A.\ Vlasenko,
{\it Phys. Rev.} C {\bf 87}, 044304 (2013).

\bibitem{BMR_PS2015} J.P.\ Blocki, A.G.\ Magner, and P.\ Ring,
{\it Phys. Scr.} T {\bf 90}, 114009 (2015).

\bibitem{BMR_PRC2015} J.P.\ Blocki, A.G.\ Magner, and P.\ Ring, 
{\it Phys. Rev.} {\bf C92}, 064311 (2015). 

\bibitem{byrd} P.F.\ Byrd and M.D.\ Friedman, 
{\it Handbook of Elliptic Integrals
   for Engineers and Scientists} (Springer-Verlag, New York, 1971).


\end{thebibliography}
\end{document}